\documentclass{amsart}

\usepackage[T1]{fontenc}
\usepackage[utf8]{inputenc}
\usepackage[hidelinks]{hyperref}
\usepackage{times}
\usepackage{xypic}

\newcommand{\NN}{\mathbb{N}} 
\newcommand{\RR}{\mathbb{R}} 
\newcommand{\ZZ}{\mathbb{Z}} 

\newcommand{\theory}[1]{\mathsf{#1}} 
\newcommand{\signature}[1]{\Sigma_{\theory{#1}}} 
\newcommand{\equations}[1]{\mathcal{E}_{\theory{#1}}} 

\newcommand{\Mod}[1]{\text{\textbf{Mod}}(\theory{#1})} 
\newcommand{\ModC}[2]{\text{\textbf{Mod}}_{#1}(\theory{#2})} 
\newcommand{\ComodC}[2]{\text{\textbf{Comod}}_{#1}(\theory{#2})} 

\newcommand{\category}[1]{\text{\textbf{#1}}} 
\newcommand{\opcat}[1]{{#1}^{\mathsf{op}}} 

\newcommand{\carrier}[1]{|#1|} 
\newcommand{\Free}[2]{F_{\theory{#1}}(#2)} 
\newcommand{\FreeFun}[1]{F_{\theory{#1}}} 

\newcommand{\all}[1]{\forall #1 \,.\,} 
\newcommand{\some}[1]{\exists #1 \,.\,} 
\newcommand{\set}[1]{\{#1\}} 

\newcommand{\lam}[1]{\lambda #1 \,.\,}

\newcommand{\family}[2]{\{#1\}_{#2}} 

\newcommand{\finpow}[1]{\mathcal{P}_{{<}\omega}(#1)} 

\newcommand{\Tree}[2]{\mathsf{Tree}_{#1}(#2)} 
\newcommand{\leaf}[1]{\return{#1}} 

\newcommand{\op}[1]{\mathsf{op}_{#1}} 
\newcommand{\arity}[1]{\mathsf{ar}_{#1}} 
\newcommand{\opdecl}[3]{#1 : #2 \leadsto #3} 

\newcommand{\one}{\mathsf{1}} 
\newcommand{\unit}{{\text{\small$()$}}} 

\newcommand{\lift}[1]{#1^\dagger} 

\newcommand{\Cinfty}{\mathcal{C}^\infty}

\newcommand{\sem}[1]{[\![#1]\!]} 

\newcommand{\bool}{\mathsf{bool}} 
\newcommand{\true}{\mathsf{true}}
\newcommand{\false}{\mathsf{false}}
\newcommand{\cond}[3]{\mathsf{if}\;#1\;\mathsf{then}\;#2\;\mathsf{else}\;#3}

\newcommand{\hto}{\Rightarrow} 

\newcommand{\defeq}{\mathbin{{:}{=}}} 

\newcommand{\tensor}[2]{#1 \otimes #2} 

\newcommand{\ct}[1]{\underline{#1}} 
\newcommand{\dirt}[2]{#1 \mathbin{!} #2} 

\newcommand{\kode}[1]{\mathsf{#1}}

\newcommand{\seq}[2]{\kode{do}\; #1 \leftarrow #2 \;\kode{in}\;}
\newcommand{\handler}{\kode{handler}\;}
\newcommand{\xopgen}[2]{\overline{#1}(#2)}
\newcommand{\opgen}[2]{\xopgen{\kode{#1}}{#2}}
\newcommand{\opcall}[3]{\kode{#1}(#2, #3)}
\newcommand{\opclause}[3]{#1(#2; #3) \mapsto}
\newcommand{\return}[1]{\kode{return}\;#1}
\newcommand{\retclause}[1]{\return{#1} \mapsto}
\newcommand{\withhandle}[2]{\kode{with}\; #1\; \kode{handle}\; #2}

{\theoremstyle{definition}
\newtheorem{example}{Example}[section]
}

\begin{document}

\title{What is algebraic about algebraic effects and handlers?}

\author{Andrej Bauer}
\address{Andrej Bauer\\
Faculty of mathematics and Physics\\
University of Ljubljana\\
Jadranska 19\\
1000 Ljubljana\\
Slovenia}
\email{Andrej.Bauer@andrej.com}
\thanks{This material is based upon work supported by the Air Force Office of
  Scientific Research under award number FA9550-17-1-0326.}

\maketitle

This note recapitulates and expands the contents of a tutorial on the
mathematical theory of algebraic effects and handlers which I gave at the
Dagstuhl seminar \emph{``Algebraic effect handlers go
  mainstream''}~\cite{chandrasekaran18:_algeb}. It is targeted roughly at the
level of a doctoral student with some amount of mathematical training, or at
anyone already familiar with algebraic effects and handlers as programming
concepts who would like to know what they have to do with algebra.

Our goal is to draw an uninterrupted line of thought between algebra and
computational effects. We begin on the mathematical side of things, by reviewing
the classic notions of universal algebra: signatures, algebraic theories, and
their models. We then generalize and adapt the theory so that it applies to
computational effects. In the last step we replace traditional mathematical
notation with one that is closer to programming languages.

\subsubsection*{Acknowledgment}

I thank Matija Pretnar for discussion and planning of the Dagstuhl tutorial and
these notes. Section~\ref{sec:what-coalg-about} on comodels was written jointly
with Matija and is based on his Dagstuhl tutorial with the same title.

\section{Algebraic theories}
\label{sec:algebraic-theories}

In algebra we study mathematical structures that are equipped with operations satisfying
equational laws. For example, a group is a structure $(G, \mathsf{u}, {\cdot}, {}^{-1})$,
where $\mathsf{u}$~is a constant, $\cdot$~is a binary operation, and ${}^{-1}$~is a unary
operation, satisfying the familiar group identities:
\begin{gather*}
  (x \cdot y) \cdot z = x \cdot (y \cdot z),\\
  \mathsf{u} \cdot x = x = x \cdot \mathsf{u},\\
  x \cdot x^{-1} = \mathsf{u} = x^{-1} \cdot x.
\end{gather*}
There are alternative axiomatizations, for instance: a group is a monoid
$(G, \mathsf{u}, {\cdot})$ in which every element is invertible, i.e.,
$\all{x \in G} \some{y \in G} x \cdot y = \mathsf{u} = y \cdot x$. However, a
formulation all of whose axioms are equations is preferred, because its simple
logical form grants its models good structural properties.

It is important to distinguish the theory of an algebraic structure from the
algebraic structures it describes. In this section we shall study the
descriptions, which are known as \emph{algebraic} or \emph{equational theories}.

\subsection{Signatures, terms and equations}
\label{sec:signatures-equations}

A \emph{signature $\Sigma$} is a collection of \emph{operation symbols} with
\emph{arities} $\family{(\op{i}, \arity{i})}{i}$. The operation symbols
$\op{i}$ may be any anything, but are usually thought of as syntactic entities,
while arities $\arity{i}$ are non-negative integers. An operation symbol whose
arity is~$0$ is called a \emph{constant} or a \emph{nullary} symbol. Operation
symbols with arities $1$, $2$ and $3$ are referred to as \emph{unary},
\emph{binary}, and \emph{ternary}, respectively.

A (possibly empty) list of distinct variables $x_1, \ldots, x_k$ is called a
\emph{context}. The \emph{$\Sigma$-terms in context $x_1, \ldots, x_k$} are
built inductively using the following rules:
\begin{enumerate}
\item each variable $x_i$ is a $\Sigma$-term in context $x_1, \ldots, x_k$,
\item if $t_1, \ldots, t_{\arity{i}}$ are $\Sigma$-terms in context $x_1, \ldots, x_k$ then
  $\op{i}(t_1, \ldots, t_{\arity{i}})$ is a $\Sigma$-term in context $x_1, \ldots, x_k$.
\end{enumerate}
We write
\begin{equation*}
  x_1, \ldots, x_k \mid t
\end{equation*}
to indicate that $t$ is a $\Sigma$-term in the given context. A \emph{closed
  $\Sigma$-term} is a $\Sigma$-term in the empty context. No variables occur
in a closed term.

A \emph{$\Sigma$-equation} is a pair of $\Sigma$-terms $\ell$ and $r$ in a context
$x_1, \ldots, x_k$. We write
\begin{equation*}
  x_1, \ldots, x_k \mid \ell = r
\end{equation*}
to indicate an equation in a context. We shall often elide the context and write simply
$\ell = r$, but it should be understood that there is an ambient context which contains at
least all the variables mentioned by $\ell$ and $r$.

A $\Sigma$-equation really is just a list of variables and a pair of terms, and
\emph{not} a logical statement. The context variables are \emph{not} universally
quantified, and we are not talking about first-order logic. Of course, a
$\Sigma$-equation is suggestively written as an equation because we do
eventually want to \emph{interpret} it as an assertion of equality, but until
such time (and even afterwards) it is better to think of contexts, terms, and
equations as ordinary mathematical objects, devoid of any imagined or special
meta-mathematical status. This remark will hopefully become clearer in
Section~\ref{sec:algebr-theor-with}.

When no confusion can arise we drop the prefix ``$\Sigma$-'' and simply speak about
terms and equations instead of $\Sigma$-terms and $\Sigma$-equations.

\begin{example}
  \label{ex:monoid-signature}
  The signature for the theory of a monoid has a nullary symbol $\mathsf{u}$ and a binary
  symbol $\mathsf{m}$. There are infinitely many expressions in context $x, y$, such as
  \begin{align*}
    \mathsf{u}(),\quad
    x,\quad
    y,\quad
    \mathsf{m}(\mathsf{u}(), \mathsf{u}()),\quad
    \mathsf{m}(\mathsf{u}(), x),\quad
    \mathsf{m}(y, \mathsf{u}()),\quad
    \mathsf{m}(x, x),\quad
    \mathsf{m}(y, x),
    \ldots
  \end{align*}
  An equation in context $x, y$ is
  \begin{equation*}
    x, y \mid \mathsf{m}(y, x) = \mathsf{m}(\mathsf{m}(\mathsf{u}(), x), y).
  \end{equation*}
  It is customary to write a nullary symbol $\mathsf{u}()$ simply as $\mathsf{u}$, and to
  use the infix operator~$\cdot$ in place of~$\mathsf{m}$. With such notation the
  above equation would be written as
  \begin{equation*}
    x, y \mid y \cdot x = (\mathsf{u} \cdot x) \cdot y.
  \end{equation*}
  One might even omit $\cdot$ and the context, in which case the equation is
  written simply as $y \, x = (\mathsf{u} \, x) \, y$. If we agree that $\cdot$
  associates to the left then $(\mathsf{u} \, x) \, y$ may be written as
  $\mathsf{u} \, x \, y$, and we are left with $y \, x = \mathsf{u} \, x \, y$,
  which is what your algebra professor might write down. Note that we are
  \emph{not} discussing validity of equations but only ways of displaying them.
\end{example}

\subsection{Algebraic theories}
\label{sec:algebraic-theories-1}

An \emph{algebraic theory $\theory{T} = (\signature{T}, \equations{T})$}, also
called an \emph{equational theory}, is given by a signature~$\signature{T}$ and
a collection $\equations{T}$ of $\signature{T}$-equations.
We impose no restrictions on the number of operation symbols or equations, but at least in
classical treatments of the subject certain complications are avoided by insisting that
arities be non-negative integers.

\begin{example}
  \label{ex:theory-group}
  The theory~$\theory{Group}$ of a group is algebraic. In order to follow closely the
  definitions we eschew the traditional notation $\cdot$ and ${}^{-1}$, and explicitly
  display the contexts. We abide by such formalistic requirements once to demonstrate
  them, but shall take notational liberties subsequently.
  The signature $\signature{Group}$ is given by operation symbols $\mathsf{u}$,
  $\mathsf{m}$, and $\mathsf{i}$ whose arities are $0$, $2$, and $1$, respectively. The
  equations $\equations{Group}$ are:
  \begin{align*}
    x, y, z &\mid \mathsf{m}(\mathsf{m}(x, y), z) = \mathsf{m}(x, \mathsf{m}(y, z)),\\
    x &\mid \mathsf{m}(\mathsf{u}(), x) = x \\
    x &\mid \mathsf{m}(x, \mathsf{u}()) = x,\\
    x &\mid \mathsf{m}(x, \mathsf{i}(x)) = \mathsf{u}()\\
    x &\mid \mathsf{m}(\mathsf{i}(x), x) = \mathsf{u}().
  \end{align*}
\end{example}

\begin{example}
  \label{ex:semi-lattice}
  The theory $\theory{Semilattice}$ of a semilattice is algebraic. It is given by a
  nullary symbol $\bot$ and a binary symbol $\vee$, satisfying the equations
  \begin{align*}
    x \vee (y \vee z) &= (x \vee y) \vee z,\\
    x \vee y &= y \vee x,\\
    x \vee x &= x,\\
    x \vee \bot &= x.
  \end{align*}
  It should be clear that the first equation has context $x, y, z$, the second one
  in~$x, y$, and the last two in~$x$.
\end{example}

\begin{example}
  \label{ex:field}
  The theory of a field, as usually given, is not algebraic because the inverse~$0^{-1}$
  is undefined, whereas the operations of an algebraic theory are always taken to be
  total. However, a proof is required to show that there is no equivalent algebraic theory.
\end{example}

\begin{example}
  \label{ex:pointed-set}
  The theory $\theory{Set_\bullet}$ of a \emph{pointed set} has a constant $\bullet$ and
  no equations.
\end{example}

\begin{example}
  \label{ex:theory-empty}
  The \emph{empty theory $\theory{Empty}$} has no operation symbols and no equations.
\end{example}

\begin{example}
  \label{ex:theory-singleton}
  The theory of a \emph{singleton $\theory{Singleton}$} has a constant $\star$ and the
  equation $x = y$.
\end{example}

\begin{example}
  \label{ex:lattice}
  A bounded lattice is a partial order with finite infima and suprema. Such a formulation
  is not algebraic because the infimum and supremum operators do not have fixed arities,
  but we can reformulate it in terms of nullary and binary operations. Thus, the theory
  $\theory{Lattice}$ of a bounded lattice has constants $\bot$ and $\top$, and two binary
  operation symbols $\vee$ and $\wedge$, satisfying the equations:
  \begin{align*}
    x \vee (y \vee z) &= (x \vee y) \vee z,   &      x \wedge (y \wedge z) &= (x \wedge y) \wedge z,\\
    x \vee y &= y \vee x,                     &      x \wedge y &= y \wedge x,\\
    x \vee x &= x,                            &      x \wedge x &= x,\\
    x \vee \bot &= x,                         &      x \wedge \top &= x.
   \intertext{In addition we need the \emph{absorbtion} laws:}
    x \vee (x \wedge y) &= x,                  & x \wedge (x \vee y) &= x.
  \end{align*}
  Notice that the theory of a bounded lattice is the juxtaposition of two copies of
  the theory of a semi-lattice from Example~\ref{ex:semi-lattice}, augmented with laws that relate them.
  The partial order is recovered because $x \leq y$ is equivalent to
  $x \vee y = y$ and to $x \wedge y = x$.
\end{example}

\begin{example}
  \label{ex:finitely-generated-group}
  A \emph{finitely generated group} is a group which contains a finite collection of
  elements, called the \emph{generators}, such that every element of the group is obtained
  by multiplications and inverses of the generators. It is not clear how to express this
  condition using only equations, but a proof is required to show that there is no
  equivalent algebraic theory.
\end{example}

\begin{example}
  \label{ex:Cinfty-theory}
  An example of an algebraic theory with many operations and equations is the theory of a
  $\Cinfty$-ring. Let $\Cinfty(\RR^n, \RR^m)$ be the set of all smooth maps from $\RR^n$
  to $\RR^m$. The signature for the theory of a $\Cinfty$-ring contains an $n$-ary
  operation symbol $\op{f}$ for each $f \in \Cinfty(\RR^n, \RR)$. For all
  $f \in \Cinfty(\RR^n, \RR)$, $h \in \Cinfty(\RR^m, \RR)$, and
  $g_1, \ldots, g_n \in \Cinfty(\RR^m, \RR)$ such that
  \begin{equation*}
    f \circ (g_1, \ldots, g_n) = h,
  \end{equation*}
  the theory has the equation
  \begin{equation*}
    x_1, \ldots, x_m \mid
    \op{f} (\op{g_1}(x_1, \ldots, x_m), \ldots, \op{g_n}(x_1, \ldots, x_m)) =
    \op{h}(x_1, \ldots, x_m).
  \end{equation*}
  The theory contains the theory of a commutative unital ring as a subtheory. Indeed,
  the ring operations on~$\RR$ are smooth maps, and so they appear as $\op{+}$,
  $\op{\times}$, $\op{-}$ in the signature, and so do constants $\op{0}$ and $\op{1}$,
  because all maps $\RR^0 \to \RR$ are smooth. The commutative ring equations are present
  as well because the real numbers form a commutative ring.
\end{example}

\subsection{Interpretations of signatures}
\label{sec:interp-of-sign}

Let a signature $\Sigma$ be given. An \emph{interpretation~$I$} of $\Sigma$ is given by
the following data:
\begin{enumerate}
\item a set $\carrier{I}$, called the \emph{carrier},
\item for each operation symbol $\op{i}$ a map
  \begin{equation*}
    \sem{\op{i}}_I : \underbrace{\carrier{I} \times \cdots \times \carrier{I}}_{\arity{i}} \to \carrier{I},
  \end{equation*}
  called an \emph{operation}.
\end{enumerate}
The double bracket $\sem{{\ }}_I$ is called the \emph{semantic bracket} and is typically
used when syntactic entities (operation symbols, terms, equations) are mapped to
their mathematical counterparts. When no confusion can arise, we omit the subscript~$I$
and write just~$\sem{{\ }}$.

We abbreviate an $n$-ary product $\carrier{I} \times \cdots \times \carrier{I}$ as $\carrier{I}^n$. A
nullary product $\carrier{I}^0$ contains a single element, namely the empty tuple
$\unit$, so it makes sense to write $\carrier{I}^0 = \one = \set{\unit}$. Thus a nullary
operation symbol is interpreted by an map $\one \to \carrier{I}$, and such maps are in
bijective correspondence with the elements of~$\carrier{I}$, which would be the constants.

An interpretation~$I$ may be extended to $\Sigma$-terms. A $\Sigma$-term in context
\begin{equation*}
  x_1, \ldots, x_k \mid t
\end{equation*}
is interpreted by a map
\begin{equation*}
  \sem{x_1, \ldots, x_k \mid t}_I : \carrier{I}^k \to \carrier{I},
\end{equation*}
as follows:
\begin{enumerate}
\item the variable $x_i$ is interpreted as the $i$-th projection,
  \begin{equation*}
    \sem{x_1, \ldots, x_k \mid  x_i}_I = \pi_i : \carrier{I}^k \to \carrier{I},
  \end{equation*}
\item a compound term in context
  \begin{equation*}
    x_1, \ldots, x_k \mid \op{i}(t_1, \ldots, t_{\arity{i}})
  \end{equation*}
  is interpreted as the composition of maps
  \begin{equation*}
    \xymatrix@+6em{
      {\carrier{I}^k} \ar[r]^{(\sem{t_1}_I, \ldots, \sem{t_{\arity{i}}}_I)}
      &
      {\carrier{I}^{\arity{i}}} \ar[r]^{\sem{\op{i}}_I}
      &
      {\carrier{I}}
    }
  \end{equation*}
  where we elided the contexts $x_1, \ldots, x_k$ for the sake of brevity.
\end{enumerate}

\begin{example}
  One interpretation of the signature from Example~\ref{ex:monoid-signature} is given by
  the carrier set $\RR$ and the interpretations of operation symbols
  \begin{align*}
    \sem{\mathsf{u}}() &= 1 + \sqrt{5}, \\
    \sem{\mathsf{m}}(a, b) &= a^2 + b^3.
  \end{align*}
  The term in context $x, y \mid \mathsf{m}(\mathsf{u}, \mathsf{m}(x, x))$ is interpreted
  as the map $\RR \times \RR \to \RR$, given by the rule
  \begin{equation*}
    (a, b) \mapsto (a+1)^3 a^6 + 2 (3 + \sqrt{5}).
  \end{equation*}
  The same term in a context $y, x, z$ is interpreted as the map $\RR \times \RR \times \RR \to \RR$,
  given by the rule
  \begin{align*}
    (a, b, c) &\mapsto (b+1)^3 b^6 + 2 (3 + \sqrt{5}).
  \end{align*}
  These are not the same map, as they do not even have the same domains!
\end{example}

The previous examples shows why contexts should not be ignored. In mathematical practice
contexts are often relegated to guesswork for the reader, or are handled implicitly. For
example, in real algebraic geometry the solution set of the equation $x^2 + y^2 = 1$ is
either a unit circle in the plane or an infinitely extending cylinder of unit radius in
the space, depending on whether the context might be $x, y$ or $x, y, z$. Which context is
meant is indicated one way or another by the author of the mathematical text.

\subsection{Models of algebraic theories}
\label{sec:models-algebr-theor}

A \emph{model~$M$} of an algebraic theory~$\theory{T}$ is an interpretation of the signature
$\signature{T}$ which validates all the equations $\equations{T}$. That is, for every
equation
\begin{equation*}
  x_1, \ldots, x_k \mid \ell = r
\end{equation*}
in~$\equations{T}$, the maps
\begin{equation*}
  \sem{x_1, \ldots, x_k \mid \ell}_M : \carrier{M}^k \to \carrier{M}
  \qquad\text{and}\qquad
  \sem{x_1, \ldots, x_k \mid r}_M : \carrier{M}^k \to \carrier{M}
\end{equation*}
are equal. We refer to a model of~$\theory{T}$ as a \emph{$\theory{T}$-model} or
a \emph{$\theory{T}$-algebra}.

\begin{example}
  A model $G$ of $\theory{Group}$, cf.\ Example~\ref{ex:theory-group}, is given by a
  carrier set $\carrier{G}$ and maps
  \begin{equation*}
    \sem{u}_G : \one \to \carrier{G},\qquad
    \sem{m}_G : \carrier{G} \times \carrier{G} \to \carrier{G},\qquad
    \sem{i}_G : \carrier{G} \to \carrier{G},
  \end{equation*}
  interpreting the operation symbols $\mathsf{u}$, $\mathsf{m}$, and
  $\mathsf{i}$, respectively, such that the equations~$\equations{Group}$. This
  amounts precisely to $(\carrier{G}, \sem{u}_G, \sem{m}_G, \sem{i}_G)$ being a
  group, except that the unit is viewed as a map $\one \to \carrier{G}$ instead
  of an element of~$\carrier{G}$.
\end{example}

\begin{example}
  Every algebraic theory has the \emph{trivial model}, whose carrier is the
  singleton~$\one$, and whose operations are interpreted by the unique maps
  $\one^k \to \one$. All equations are satisfied because any two maps $\one^k \to \one$
  are equal.
\end{example}

The previous example explains why one should \emph{not} require $0 \neq 1$ in a ring, as
that prevents the theory of a ring from being algebraic.

\begin{example}
  The empty set is a model of a theory~$\theory{T}$ if, and only if, every operation symbol
  of $\theory{T}$ has non-zero arity.
\end{example}

\begin{example}
  A model of the theory~$\theory{Set_\bullet}$ of a pointed set, cf.\
  Example~\ref{ex:pointed-set}, is a set $S$ together with an element $s \in S$ which
  interprets the constant~$\bullet$.
\end{example}

\begin{example}
  A model of the theory~$\theory{Empty}$, cf.\ Example~\ref{ex:theory-empty}, is the same
  thing as a set.
\end{example}

\begin{example}
  A model of the theory~$\theory{Singleton}$, cf.\ Example~\ref{ex:theory-singleton}, is
  any set with precisely one element.
\end{example}

Suppose $L$ and $M$ are models of a theory~$\theory{T}$. Then we may form the
\emph{product of models} $L \times M$ by taking the cartesian product as the carrier,
\begin{equation*}
  \carrier{L \times M} = \carrier{L} \times \carrier{M},
\end{equation*}
and pointwise operations,
\begin{equation*}
  \sem{\op{i}}_{M \times L}(a, b) = (\sem{\op{i}}_M(a), \sem{\op{i}}_L(b)).
\end{equation*}
The equations $\equations{T}$ are valid in $L \times M$ because they are valid on each
coordinate separately. This construction can be extended to a product of any number of
models, including an infinite one.

\begin{example}
  We may now prove that the theory of a field from Example~\ref{ex:field} is not
  equivalent to an algebraic theory. There are fields of size 2 and 3, namely $\ZZ_2$ and
  $\ZZ_3$. If there were an algebraic theory of a field, then $\ZZ_2 \times \ZZ_3$ would
  be a field too, but it is not, and in fact there is no field of size~6.
\end{example}

\begin{example}
  Similarly, the theory of a finitely generated group from
  Example~\ref{ex:finitely-generated-group} cannot be formulated as an algebraic theory,
  because an infinite product of non-trivial finitely generated groups is not finitely
  generated.
\end{example}

\begin{example}
  Let us give a model of the theory of a $\Cinfty$-ring from
  Example~\ref{ex:Cinfty-theory}. Pick a smooth manifold~$M$, and let the carrier be the
  set $\Cinfty(M, \RR)$ of all smooth scalar fields on~$M$. Given
  $f \in \Cinfty(\RR^n, \RR)$, interpret the operation $\op{f}$ as composition with~$f$,
  \begin{align*}
    \sem{\op{f}} &: \Cinfty(M, \RR)^n \to \Cinfty(M, \RR) \\
    \sem{\op{f}} &: (u_1, \ldots, u_n) \mapsto f \circ (u_1, \ldots, u_n).
  \end{align*}
  We leave it as an exercise to verify that all equations are validated by this
  interpretation.
\end{example}

\subsection{Homomorphisms and the category of models}
\label{sec:homom-categ-models}

Suppose $L$ and $M$ are models of a theory~$\theory{T}$. A
\emph{$\theory{T}$-homomorphism} from $L$ to $M$ is a map $\phi : \carrier{L} \to \carrier{M}$ between the
carriers which commutes with operations: for every operation symbol $\op{i}$
of~$\theory{T}$, we have
\begin{equation*}
  \phi \circ \sem{\op{i}}_L = \sem{\op{i}}_M \circ \underbrace{(\phi, \ldots, \phi)}_{\arity{i}}.
\end{equation*}

\begin{example}
  A homomorphism between groups $G$ and $H$ is a map $\phi : \carrier{G} \to \carrier{H}$ between the
  carriers such that, for all $a, b \in \carrier{G}$,
  \begin{align*}
    \phi(\sem{\mathsf{u}()}_G) &= \sem{\mathsf{u}()}_H,\\
    \phi(\sem{\mathsf{m}}_G (a,b)) &= \sem{\mathsf{m}}_H (\phi(a), \phi(b)),\\
    \phi(\sem{\mathsf{i}}_G (a)) &= \sem{\mathsf{i}}_H (\phi(a)).
  \end{align*}
  This is a convoluted way of saying that the unit maps to the unit, and that
  $\phi$ commutes with the group operation and the inverses. Algebra textbooks
  usually require only that a group homomorphism commute with the group
  operation, which then implies that it also preserves the unit and commutes
  with the inverse.
\end{example}

We may organize the models of an algebraic theory~$\theory{T}$ into a category $\Mod{T}$
whose objects are the models of the theory, and whose morphisms are homomorphisms of the
theory.

\begin{example}
  The category of models of theory $\theory{Group}$, cf.\ Example~\ref{ex:theory-group},
  is the usual category of groups and group homomorphisms.
\end{example}

\begin{example}
  The category of models of the theory $\theory{Set_\bullet}$, cf.\
  Example~\ref{ex:pointed-set}, has as its objects the pointed sets, which are pairs
  $(S, s)$ with $S$ a set and $s \in S$ its \emph{point}, and as homomorphisms
  the point-preserving functions between sets.
\end{example}

\begin{example}
  The category of models of the empty theory $\theory{Empty}$, cf.\
  Example~\ref{ex:theory-empty}, is just the category $\category{Set}$ of sets and
  functions.
\end{example}

\begin{example}
  The category of models of the theory of a singleton $\theory{Singleton}$, cf.\
  Example~\ref{ex:theory-singleton}, is the category whose objects are all the singleton
  sets. There is precisely one morphisms between any two of them. This category is
  equivalent to the trivial category which has just one object and one morphism.
\end{example}

\subsection{Models in a category}
\label{sec:models-category}

So far we have taken the models of an algebraic theory to be sets. More generally, we may
consider models in any category $\category{C}$ with finite products. Indeed, the
definitions of an interpretation and a model from Sections~\ref{sec:interp-of-sign}
and~\ref{sec:models-algebr-theor} may be directly transcribed so that they apply
to~$\category{C}$. An interpretation $I$ in $\category{C}$ is given by
\begin{enumerate}
\item an object $\carrier{I}$ in $\category{C}$, called the \emph{carrier},
\item for each operation symbol $\op{i}$ a morphism in $\category{C}$
  \begin{equation*}
    \sem{\op{i}}_I : \underbrace{\carrier{I} \times \cdots \times \carrier{I}}_{\arity{i}} \to \carrier{I}.
  \end{equation*}
\end{enumerate}
Once again, we abbreviate the $k$-fold product of $\carrier{I}$ as $\carrier{I}^k$. Notice that a nullary
symbol is interpreted as a morphism $\carrier{I}^0 \to \carrier{I}$, which is a morphisms from the
terminal object $\one \to \carrier{I}$ in~$\category{C}$.

An interpretation~$I$ is extended to $\Sigma$-terms in contexts as follows:
\begin{enumerate}
\item the variable $x_1, \ldots, x_k \mid x_i$ is interpreted as the $i$-th projection,
  \begin{equation*}
    \sem{x_1, \ldots, x_k \mid  x_i}_I = \pi_i : \carrier{I}^k \to \carrier{I},
  \end{equation*}
\item a compound term in context
  \begin{equation*}
    x_1, \ldots, x_k \mid \op{i}(t_1, \ldots, t_{\arity{i}})
  \end{equation*}
  is interpreted as the composition of morphisms
  \begin{equation*}
    \xymatrix@+6em{
      {\carrier{I}^k} \ar[r]^{(\sem{t_1}_I, \ldots, \sem{t_{\arity{i}}}_I)}
      &
      {\carrier{I}^{\arity{i}}} \ar[r]^{\sem{\op{i}}_I}
      &
      {\carrier{I}}
    }
  \end{equation*}
\end{enumerate}
A model of an algebraic theory~$\theory{T}$ in~$\category{C}$ is an interpretation~$M$ of
its signature $\signature{T}$ which validates all the equations. That is, for every
equation
\begin{equation*}
  x_1, \ldots, x_k \mid \ell = r
\end{equation*}
in $\equations{T}$, the morphisms
\begin{equation*}
  \sem{x_1, \ldots, x_k \mid \ell}_M : \carrier{M}^k \to \carrier{M}
  \qquad\text{and}\qquad
  \sem{x_1, \ldots, x_k \mid r}_M : \carrier{M}^k \to \carrier{M}
\end{equation*}
are equal.

The definition of a homomorphism carries over to the general setting as well. A
\emph{$\theory{T}$-homomorphism} between $\theory{T}$-models $L$ and $M$ in a category
$\category{C}$ is a morphism $\phi : \carrier{L} \to \carrier{M}$ in~$\category{C}$ such that, for every
operation symbol~$\op{i}$ in~$\theory{T}$, $\phi$ commutes with the interpretation of
$\op{i}$,
\begin{equation*}
  \phi \circ \sem{\op{i}}_L = \sem{\op{i}}_M \circ \underbrace{(\phi, \ldots, \phi)}_{\arity{i}}.
\end{equation*}
The $\theory{T}$-models and $\theory{T}$-homomorphisms in a category $\category{C}$ form a
category $\ModC{\category{C}}{T}$.

\begin{example}
  A model of the theory $\theory{Group}$ in the category $\category{Top}$ of
  topological spaces and continuous maps is a topological group.
\end{example}

\begin{example}
  What is a model of the theory $\theory{Group}$ in the category of groups $\category{Grp}$? Its
  carrier is a group $(G, u, m, i)$ together with group homomorphisms
  $\upsilon : \one \to G$, $\mu : G \times G \to G$, and $\iota : G \to G$ which satisfy
  the group laws. Because $\upsilon$ is a group homomorphism, it maps the unit of the
  trivial group~$\one$ to $u$, so the units $u$ and $\upsilon$ agree. The operations $m$
  and $\mu$ agree too, because
  \begin{equation*}
    \mu(x, y) =
    \mu(m(x, u), m(u, y)) =
    m(\mu(x, u), \mu(u, y)) =
    m(x, y),
  \end{equation*}
  where in the middle step we used the fact that $\mu$ is a group homomorphism. It is now
  clear that the inverses $i$ and $\iota$ agree as well. Furthermore, taking into account
  that $m$ and $\mu$ agree, we also obtain
  \begin{equation*}
    m(x, y) =
    m(m(u, x), m(y, u)) =
    m(m(u, y), m(x, u)) =
    m(y, x).
  \end{equation*}
  The conclusion is that a group in the category of groups is an abelian group. The
  category $\ModC{\category{Grp}}{Group}$ is therefore equivalent to the category of abelian groups.
\end{example}

\begin{example}
  A model of the theory of a pointed set, cf.\ Example~\ref{ex:pointed-set}, in the
  category of groups $\category{Grp}$ is a group $(G, u, m, i)$ together with a
  homomorphism $\one \to G$ from the trivial group~$\one$ to $G$. However, there is
  precisely one such homomorphism which therefore need not be mentioned at all. Thus a
  pointed set in groups amounts to a group.
\end{example}

\subsection{Free models}
\label{sec:free-models}

Of special interest are the free models of an algebraic theory. Given an
algebraic theory~$\theory{T}$ and a set $X$, the \emph{free $\theory{T}$-model},
also called the \emph{free $\theory{T}$-algebra}, generated by~$X$ is a model
$M$ together with a map $\eta : X \to \carrier{M}$ such that, for every
$\theory{T}$-model $L$ and every map $f : X \to \carrier{L}$ there is a unique
$\theory{T}$-homomorphism $\overline{f} : M \to L$ for which the following
diagram commutes:
\begin{equation*}
  \xymatrix{
    {X}
    \ar[r]^{\eta}
    \ar[rd]_{f}
    &
    {\carrier{M}}
    \ar[d]^{\overline{f}}
    \\
    &
     \carrier{L}
  }
\end{equation*}
The definition is a bit of a mouthful, but it can be understood as follows: the
free $\theory{T}$-model generated by $X$ is the ``most economical'' way of
making a $\theory{T}$-model out of the set~$X$.

\begin{example}
  The free group generated by the empty set is the trivial group~$\one$ with
  just one element. The map $\eta : \emptyset \to \one$ is the unique one, and
  given any (unique) map $f : \emptyset \to \carrier{G}$ to a carrier of another
  group~$G$, there is a unique group homomorphism $\overline{f} : \one \to G$.
  The relevant triangle commutes automatically because it originates
  at~$\emptyset$.
\end{example}

\begin{example}
  The free group generated by the singleton set $\one$ is the group of integers
  $(\ZZ, 0, {+}, {-})$. The map $\eta : \set{\star} \to \ZZ$ takes the generator
  $\unit$ to $0$. As an exercise you should verify that the integers have the
  required universal property.
\end{example}

\begin{example}
  Let $\finpow{X}$ be the set of all finite subsets of a set~$X$. We show that
  $(\finpow{X}, \emptyset, {\cup})$ is the free semilattice generated by~$X$, cf.\
  Example~\ref{ex:semi-lattice}. The map $\eta : X \to \finpow{X}$ takes $x \in X$ to the
  singleton set $\eta(x) = \set{x}$. Given any semilattice $(L, \bot, {\vee})$ and a map
  $f : X \to \carrier{L}$, define the homomorphism $\overline{f} : \finpow{X} \to \carrier{L}$ by
  \begin{equation*}
    \overline{f}(\set{x_1, \ldots, x_n}) = f(x_1) \wedge \cdots \wedge f(x_n).
  \end{equation*}
  Clearly, the required diagram commutes because
  \begin{equation*}
    \overline{f}(\eta(x)) = \overline{f}(\set{x}) = f(x).
  \end{equation*}
  If $g : \finpow{X} \to \carrier{L}$ is another homomorphism satisfying $g \circ \eta = f$ then
  \begin{multline*}
    g(\set{x_1, \ldots, x_n})
    = g(\eta(x_1) \cup \cdots \cup \eta(x_n))
    = g(\eta(x_1)) \wedge \cdots \wedge g(\eta(x_n)) \\
    = f(x_1) \wedge \cdots \wedge f(x_n)
    = \overline{f}(\set{x_1, \ldots, x_n}),
  \end{multline*}
  hence $\overline{f}$ is indeed unique.
\end{example}

\begin{example}
  The free model generated by~$X$ of the theory of a pointed set, cf.\
  Example~\ref{ex:pointed-set}, is the disjoint union $X + \one$ whose elements are of the
  form $\iota_1(x)$ for $x \in X$ and $\iota_2(y)$ for $y \in \one$. The point is the
  element $\iota_2(\unit)$. The map $\eta : X \to X + \one$ is the canonical
  inclusion~$\iota_1$.
\end{example}

\begin{example}
  The free model generated by~$X$ of the empty theory, cf.\ Example~\ref{ex:theory-empty},
  is~$X$ itself, with $\eta : X \to X$ the identity map.
\end{example}

\begin{example}
  The free model generated by~$X$ of the theory of a singleton, cf.\
  Example~\ref{ex:theory-singleton}, is the singleton set~$\one$, with $\eta : X \to \one$
  the only map it could be. This example shows that~$\eta$ need not be injective.
\end{example}

Every algebraic theory~$\theory{T}$ has a free model. Let us sketch its
construction. Given a signature $\Sigma$ and a set~$X$,
define~$\Tree{\Sigma}{X}$ to be the set of well-founded trees built inductively
as follows:
\begin{enumerate}
\item for each $x \in X$, there is a tree $\leaf{x} \in \Tree{\Sigma}{X}$,
\item for each operation symbol $\op{i}$ and trees
  $t_1, \ldots, t_{\arity{i}} \in \Tree{\Sigma}{X}$, there is a tree, denoted by
  $\op{i}(t_1, \ldots, t_n) \in \Tree{\Sigma}{X}$, whose root is labeled by
  $\op{i}$ and whose subtrees are $t_1, \ldots, t_{\arity{i}}$.
\end{enumerate}
By labeling the tree leaves with the keyword ``$\mathsf{return}$'' we are
anticipating their role in effectful computations, as will become clear later
on. From a purely formal point of view the choice of the label is immaterial.

The $\Sigma$-terms in context $x_1, \ldots, x_n$ are precisely the trees in
$\Tree{\Sigma}{\set{x_1, \ldots, x_n}}$, except that a variable $x_i$ is labeled
as $\leaf{x_i}$ when construed as a tree.

Suppose $x_1, \ldots, x_n \mid t$ is a $\Sigma$-term in context, and we are
given an assignment $\sigma : \set{x_1, \ldots, x_n} \to \Tree{\Sigma}{X}$ of
trees to variables. Then we may build the tree $\sigma(t)$ inductively as
follows:
\begin{enumerate}
\item $\sigma(t) = \sigma(x_i)$ if $t = x_i$,
\item $\sigma(t) = \op{i}(\sigma(t_1), \ldots, \sigma(t_n))$ if
  $t = \op{i}(t_1, \ldots, t_n)$.
\end{enumerate}
In words, the tree $\sigma(t)$ is obtained by replacing each variable~$x_i$ in~$t$ with
the corresponding tree $\sigma(x_i)$.

Given a theory~$\theory{T}$, let $\approx_\theory{T}$ be the least equivalence relation on
$\Tree{\signature{T}}{X}$ such that:
\begin{enumerate}
\item for every equation $x_1, \ldots, x_n \mid \ell = r$ in $\equations{T}$ and for every
  assignment $\sigma : \set{x_1, \ldots, x_n} \to \Tree{\signature{T}}{X}$, we have
  \begin{equation*}
    \sigma(\ell) \approx_{\theory{T}} \sigma(r).
  \end{equation*}
\item $\approx_{\theory{T}}$ is a \emph{$\signature{T}$-congruence}: for every
  operation symbol $\op{i}$ in $\signature{T}$, and for all trees
  $s_1, \ldots, s_{\arity{i}}$ and $t_1, \ldots, t_{\arity{i}}$, if
  \begin{equation*}
    s_1 \approx_{\theory{T}} t_1,
    \quad \ldots \quad,
    s_{\arity{i}} \approx_{\theory{T}} t_{\arity{i}}
  \end{equation*}
  then
  \begin{equation*}
    \op{i}(s_1, \ldots, s_{\arity{i}}) \approx_{\theory{T}}
    \op{i}(t_1, \ldots, t_{\arity{i}}).
  \end{equation*}
\end{enumerate}
Define the carrier of the free model $\Free{T}{X}$ to be the quotient set
\begin{equation*}
  \carrier{\Free{T}{X}} = \Tree{\signature{T}}{X} / {\approx_{\theory{T}}}.
\end{equation*}
Let $[t]$ be the $\approx_{\theory{T}}$-equivalence class of
$t \in \Tree{\signature{T}}{X}$. The interpretation of the operation symbol $\op{i}$ in
  $\Free{T}{X}$ is the map $\sem{\op{i}}_{\Free{T}{X}}$ defined by
\begin{equation*}
  \sem{\op{i}}_{\Free{T}{X}}([t_1], \ldots, [t_{\arity{i}}]) =
  [\op{i}(t_1, \ldots, t_{\arity{i}})].
\end{equation*}
The map $\eta_X : X \to \Free{T}{X}$ is defined by
\begin{equation*}
  \eta_X(x) = [\leaf{x}].
\end{equation*}
To see that we successfully defined a $\theory{T}$-model, and that it is freely generated
by~$X$, one has to verify a number of mostly straightforward technical details, which we
omit.

When a theory $\theory{T}$ has no equations the free models generated by~$X$ is just the
set of trees $\Tree{T}{X}$ because the relation $\approx_{\theory{T}}$ is equality.

\subsection{Operations with general arities and parameters}
\label{sec:oper-gener-arit-param}

We have so far followed the classic mathematical presentation of algebraic
theories. To get a better fit with computational effects, we need to generalize
operations in two ways.

\subsubsection{General arities}
\label{sec:general-arities}

We shall require operations that accept an arbitrary, but fixed collection of
arguments. One might expect that the correct way to do so is to allow arities to
be ordinal or cardinal numbers, as these generalize natural numbers, but that
would be a thoroughly non-computational idea. Instead, let us observe that an
$n$-ary cartesian product
\begin{equation*}
  \underbrace{X \times \cdots \times X}_{n}
\end{equation*}
is isomorphic to the exponential $X^{[n]}$, where
$[n] = \set{0, 1, \ldots, n-1}$. Recall that an exponential $B^A$ is the set of
all functions $A \to B$, and in fact we shall use the notations $B^A$ and
$A \to B$ interchangeably. If we replace $[n]$ by an arbitrary set~$A$, then we
can think of a map
\begin{equation*}
  X^A \to X
\end{equation*}
as taking $A$-many arguments. We need reasonable notation for writing down an
operation symbol applied to $A$-many arguments, where $A$ is an arbitrary set. One
might be tempted to adapt the tuple notation and write something silly, such as
\begin{equation*}
  \op{i}(\cdots t_a \cdots)_{a \in A},
\end{equation*}
but as computer scientists we know better than that. Let us use the notation that
is already provided to us by the exponentials, namely the $\lambda$-calculus. To
have $A$-many elements of a set $X$ is to have a map $\kappa : A \to X$, and
thus to apply the operation symbol $\op{i}$ to $A$-many arguments $\kappa$ we
simply write~$\op{i}(\kappa)$.

\begin{example}
  Let us rewrite the group operations in the new notation. The empty set
  $\emptyset$, the singleton $\one$, and the set of boolean values
  \begin{equation*}
    \bool = \set{\false, \true}
  \end{equation*}
  serve as arities. We use the conditional statement
  \begin{equation*}
    \cond{b}{x}{y}
  \end{equation*}
  as a synonym for what is usually written as definition by cases,
  \begin{equation*}
  \begin{cases}
      x & \text{if $b = \true$,}\\
      y & \text{if $b = \false$.}
    \end{cases}
  \end{equation*}
  Now a group is given by a carrier set $G$ together with maps
  \begin{align*}
    \mathsf{u} &: G^\emptyset \to G,\\
    \mathsf{m} &: G^\bool \to G,\\
    \mathsf{i} &: G^\one \to G,
  \end{align*}
  satisfying the usual group laws, which we ought to write down using the
  $\lambda$-notation. The associativity law is written like this:
  \begin{multline*}
    \mathsf{m}(\lam{b} \cond{b}{\mathsf{m}(\lam{c}\cond{c}{x}{y})}{z}) = \\
    \mathsf{m}(\lam{b} \cond{b}{x}{\mathsf{m}(\lam{c} \cond{c}{y}{z})}).
  \end{multline*}
  Here is the right inverse law, where $\mathsf{O}_X : \emptyset \to X$ is
  the unique map from $\emptyset$ to~$X$:
  \begin{equation*}
    \mathsf{m}(\lam{b} \cond{b}{x}{\mathsf{i}(\lam{\_}{x})}) =
    \mathsf{u}(\mathsf{O}_G).
  \end{equation*}
  The symbol $\_$ indicates that the argument of the $\lambda$-abstraction is
  ignored, i.e., that the function defined by the abstraction is constant. One
  more example might help: $x$ squared may be written as
  $\mathsf{m}(\lam{b} \cond{b}{x}{x})$ as well as $\mathsf{m}(\lam{\_} x)$.
\end{example}

Such notation is not appropriate for performing algebraic manipulations, but is
bringing us closer to the syntax of a programming language.

\subsubsection{Operations with parameters}
\label{sec:oper-with-param}

To motivate our second generalization, consider the theory of a module~$M$ over
a ring~$R$ (if you are not familiar with modules, think of the elements of $M$
as vectors and the elements of~$R$ as scalars). For it to be an algebraic
theory, we need to deal with scalar multiplication ${\cdot} : R \times M \to M$,
because it does not fit the established pattern. There are three possibilities:
\begin{enumerate}
\item We could introduce \emph{multi-sorted} algebraic theories whose operations
  take arguments from several carrier sets. The theory of a module would have
  two sorts, say $\mathsf{R}$ and $\mathsf{M}$, and scalar multiplication would
  be a binary operation of arity $(\mathsf{R}, \mathsf{M}; \mathsf{M})$. (We
  hesitate to write $\mathsf{R} \times \mathsf{M} \to \mathsf{M}$ lest the type
  theorists get useful ideas.)
\item Instead of having a single binary operation taking a scalar and a vector,
  we could have many unary operations taking a vector, one for each scalar.
\item We could view the scalar as an additional \emph{parameter} of a
  unary operation on vectors.
\end{enumerate}
The second and the third options are superficially similar, but they differ in
their treatment of parameters. In one case the parameters are part of the
indexing of the signature, while in the other they are properly part of the
algebraic theory. We shall adopt operations with parameters because they
naturally model algebraic operations that arise as computational effects.

\begin{example}
  The theory of a module over a ring~$(R, 0, {+}, {-}, {\cdot})$ has several
  operations. One of them is scalar multiplication, which is a \emph{unary}
  operation $\mathsf{mul}$ parameterized by elements of~$R$. That is, for every
  $r \in R$ and term $t$, we may form the term
  \begin{equation*}
    \mathsf{mul}(r; t),
  \end{equation*}
  which we think of as~$t$ multiplied with~$r$. The remaining operations seem
  not to be parameterized, but we can force them to be parameterized by fiat.
  Addition is a binary operation $\mathsf{add}$ parameterized by the singleton
  set~$\one$: the sum of $t_1$ and $t_2$ is written as
  \begin{equation*}
    \mathsf{add}(\unit; t_1, t_2).
  \end{equation*}
  We can use this trick in general: an operation without parameters is an
  operation taking parameters from the singleton set.
\end{example}

Note that in the previous example we mixed theories and models. We spoke about
the \emph{theory} of a module with respect to a \emph{specific ring}~$R$.

\begin{example}
  The theory of a $\Cinfty$-ring, cf.\ Example~\ref{ex:Cinfty-theory}, may be
  reformulated using parameters. For every $n \in \NN$ there is an $n$-ary
  operation symbol $\mathsf{app}_n$ whose parameter set is
  $\Cinfty(\RR^n, \RR)$. What was written as
  \begin{equation*}
    \op{f}(t_1, \ldots, t_n)
  \end{equation*}
  in Example~\ref{ex:Cinfty-theory} is now written as
  \begin{equation*}
    \mathsf{app}_n(f; t_1, \ldots, t_n).
  \end{equation*}
  If you insist on the~$\lambda$-notation, replace the tuple
  $(t_1, \ldots, t_n)$ of terms with a single function $t$ mapping from $[n]$ to
  terms, and write $\mathsf{app}_n(f; t)$.

  The operations $\mathsf{app}_n$ tell us what $\Cinfty$-rings are about: they
  are structures whose elements can feature as arguments to smooth functions. In
  contrast, an ordinary (commutative unital) ring is one whose elements can
  feature as arguments to ``finite degree'' smooth maps, i.e., the polynomials.

\end{example}

\subsection{Algebraic theories with parameterized operations and general arities}
\label{sec:algebr-theor-with}

Let us restate the definitions of signatures and algebraic operations, with the
generalizations incorporated. For simplicity we work with sets and functions,
and leave consideration of other categories for another occasion.

A \emph{signature $\Sigma$} is given by a collection of \emph{operation symbols
  $\op{i}$} with associated \emph{parameter sets $P_i$} and
\emph{arities~$A_i$}. For reasons that will become clear later, we write
\begin{equation*}
  \opdecl{\op{i}}{P_i}{A_i}
\end{equation*}
to display an operation symbol $\op{i}$ with parameter set $P_i$ and
arity~$A_i$. The symbols may be anything, although we think of them as syntactic
entities, while $P_i$'s and $A_i$'s are sets.

Arbitrary arities require an arbitrary number of variables in context. We therefore
generalize terms in contexts to \emph{well-founded trees} over~$\Sigma$
generated by a set~$X$. These form a set $\Tree{\Sigma}{X}$ whose elements are
generated inductively as follows:
\begin{enumerate}
\item for every generator $x \in S$ there is a tree $\leaf{x}$,
\item if $p \in P_i$ and $\kappa : A_i \to \Tree{\Sigma}{X}$ then
  $\op{i}(p, \kappa)$ is a tree whose root is labeled with~$\op{i}$ and whose
  $A_i$-many subtrees are given by~$\kappa$.
\end{enumerate}
The usual $\Sigma$-terms in context $x_1, \ldots, x_k$ correspond to
$\Tree{\Sigma}{\set{x_1, \ldots, x_k}}$. Or to put it differently, the elements
of $\Tree{\Sigma}{X}$ may be thought of as terms with variables~$X$. In fact, we
shall customarily refer to them as terms.

An \emph{interpretation $I$ of a signature $\Sigma$} is given by:
\begin{enumerate}
\item a carrier set $\carrier{I}$,
\item for each operation symbol $\op{i}$ with parameter set~$P_i$ and arity~$A_i$,
  a map
  \begin{equation*}
    \sem{\op{i}}_I : P_i \times \carrier{I}^{A_i} \longrightarrow \carrier{I}.
  \end{equation*}
\end{enumerate}
The interpretation $I$ may be extended to trees. A tree $t \in \Tree{\Sigma}{X}$
is interpreted as a map
\begin{equation*}
  \sem{t}_I : \carrier{I}^X \to \carrier{I}
\end{equation*}
as follows:
\begin{enumerate}
\item the tree $\leaf{x}$ is interpreted as the $x$-th projection,
  \begin{align*}
    \sem{\leaf{x}}_I : \carrier{I}^X \to \carrier{I},\\
    \sem{\leaf{x}}_I : \eta \mapsto \eta(x),
  \end{align*}
\item the tree $\op{i}(p, \kappa)$ is interpreted as the map
  \begin{align*}
    \sem{\op{i}(p, \kappa)}_I &: \carrier{I}^X \longrightarrow \carrier{I} \\
    \sem{\op{i}(p, \kappa)}_I &:
      \eta \mapsto
      \sem{\op{i}}_I(p, \lam{a} \sem{\kappa(a)}_I(\eta)),
  \end{align*}
\end{enumerate}

A \emph{$\Sigma$-equation} is a set $X$ and a pair of $\Sigma$-terms
$\ell, r \in \Tree{\Sigma}{X}$, written
\begin{equation*}
  X \mid \ell = r.
\end{equation*}
We usually leave out~$X$. Given an interpretation $I$ of signature $\Sigma$, we
say that such an equation is \emph{valid} for~$I$ when the interpretations of
$\ell$ and $r$ give the same map.

An \emph{algebraic theory $\theory{T} = (\signature{T}, \equations{T})$} is
given by a signature $\signature{T}$ and a collection of $\Sigma$-equations
$\equations{T}$. A \emph{$\theory{T}$-model} is an interpretation for
$\signature{T}$ which validates all the equations~$\equations{T}$.

The notions of $\theory{T}$-morphisms and the category $\Mod{T}$ of
$\theory{T}$-models and $\theory{T}$-morphisms may be similarly generalized. We
do not repeat the definitions here, as they are almost the same. You should
convince yourself that every algebraic theory has a free model, which is still
built as a quotient of the set of well-founded trees.

\section{Computational effects as algebraic operations}
\label{sec:comp-effects-as}

It is high time we provide some examples from programming. The original insight
by Gordon Plotkin and John
Power~\cite{plotkin01:_seman_algeb_operat,plotkin03:_algeb_operat_gener_effec}
was that many computational effects are naturally described by algebraic
theories. What precisely does this mean?

When a program runs on a computer, it interacts with the environment by
performing \emph{operations}, such as printing on the screen, reading from the
keyboard, inspecting and modifying external memory store, launching missiles,
etc. We may model these phenomena mathematically as operations on an algebra
whose elements are \emph{computations}. Leaving the exact nature of computations
aside momentarily, we note that a computation may be
\begin{itemize}
\item pure, in which case it terminates and returns a value, or
\item effectful, in which case it performs an operation.
\end{itemize}
(We are ignoring a third possibility, non-termination.) Let us write
\begin{equation*}
  \return{v}
\end{equation*}
for a pure computation that returns the value~$v$. Think of a value as an inert
datum that needs no further computation, such as a boolean constant, a numeral,
or a $\lambda$-abstraction. An operation takes a \emph{parameter}~$p$, for
instance the memory location to be read, or the string to be printed, and a
\emph{continuation}~$\kappa$, which is a suspended computation expecting the
result of the operation, for instance the contents of the memory location that
has been read. Thus it makes sense to write
\begin{equation*}
  \opcall{op}{p}{\kappa}
\end{equation*}
for the computation that performs the operation~$\kode{op}$, with parameter~$p$
and continuation~$\kappa$. The similarity with algebraic operations from
Section~\ref{sec:algebr-theor-with} is not incidental!

\begin{example}
  The computation which increases the contents of memory location~$\ell$ by~$1$
  and returns the original contents is written as
  \begin{equation*}
    \opcall{lookup}{\ell}{
    \lam{x} \opcall{update}{(\ell,x + 1)}{
    \lam{\_} \return{x}
    }
    }.
  \end{equation*}
  In some venerable programming languages we would write this as $\ell{+}{+}$.
  Note that the operations happen from outside in: first the memory
  location~$\ell$ is read, its value is bound to~$x$, then $x + 1$ is written to
  memory location~$\ell$, the result of writing is ignored, and finally the
  value of~$x$ is returned.
\end{example}

So far we have a notation that looks like algebraic operations, but to do things
properly we need signatures and equations. These depend on the computational
effects under consideration.

\begin{example}
  \label{ex:theory-state}
  The algebraic theory of \emph{state} with \emph{locations $L$} and
  \emph{states $S$} has operations
  \begin{equation*}
    \opdecl{\kode{lookup}}{L}{S}
    \qquad\text{and}\qquad
    \opdecl{\kode{update}}{L \times S}{\one}.
  \end{equation*}
  First we have equations which state what happens on successive lookups and
  updates to the same memory location. For all $\ell \in L$, $s \in S$ and
  all continuations~$\kappa$:
  \begin{align*}
    \opcall{lookup}{\ell}{
      \lam{s}{
        \opcall{lookup}{\ell}{
        \lam{t} \kappa \, s \, t}
      }
    } &=
    \opcall{lookup}{\ell}{\lam{s}{\kappa \, s \, s}}
    \\
    \opcall{lookup}{\ell}{
      \lam{s} \opcall{update}{(\ell, s)}{\kappa}
    } &=
    \kappa \, ()
    \\
    \opcall{update}{(\ell, s)}{
      \lam{\_} \opcall{lookup}{\ell}{\kappa}
    } &=
    \opcall{update}{(\ell, s)}{\lam{\_} \kappa \, s}
    \\
    \opcall{update}{(\ell, s)}{
      \lam{\_} \opcall{update}{(\ell, t)}{\kappa}
    } &=
    \opcall{update}{(\ell, t)}{\kappa}
  \end{align*}
  For example, the first equations says that two consecutive lookups from a
  memory location give equal results. We ought to explain the precise nature
  of~$\kappa$ in the above equations. If we translate the earlier examples into
  the present notation, we see that~$\kappa$ corresponds to variables, which
  leads to the idea that we should use a \emph{generic}~$\kappa$. Thus we let
  $\kappa$ take some arguments and just return them as a tuple. In the first
  equation we take $\kappa = \lam{s \, t } \leaf {(s, t)}$; in the second and
  fourth equations are $\kappa = \lam{\_} \leaf{\unit}$; and in the third
  equation $\kappa = \lam{s} \leaf{s}$. All equations have empty contexts, as no
  free variables occur in them. Unless specified otherwise, we shall
  always take~$\kappa$ to be such a generic continuation.

  There is a second set of equations stating that lookups and updates from
  \emph{different} locations $\ell \neq \ell'$ distribute over each other:
  \begin{align*}
    \opcall{lookup}{\ell}{
       \lam{s} \opcall{lookup}{\ell'}{\lam{s'} \kappa \, s \, s'}
    } &=
    \opcall{lookup}{\ell'}{
       \lam{s'} \opcall{lookup}{\ell}{\lam{s} \kappa \, s \, s'}
    }
    \\
    \opcall{update}{(\ell, s)}{
       \lam{\_} \opcall{lookup}{\ell'}{\kappa}
    } &=
    \opcall{lookup}{\ell'}{
       \lam{t} \opcall{update}{(\ell, s)}{
          \lam{\_} \kappa \, t
       }
    } \\
    \opcall{update}{(\ell, s)}{
       \lam{\_} \opcall{update}{(\ell', s')}{\kappa}
    } &=
    \opcall{update}{(\ell', s')}{
       \lam{\_} \opcall{update}{(\ell, s)}{\kappa}
    }.
  \end{align*}
  Have we forgotten any equations?
  It turns out that the theory is Hilbert-Post complete: if we add any equation
  that does not already follow from these, the theory trivializes in the sense
  that all equations become derivable.
\end{example}

\begin{example}
  \label{ex:theory-io}
  The theory of \emph{input and output} (I/O) has operations
  \begin{equation*}
    \opdecl{\kode{print}}{S}{\one}
    \qquad\text{and}\qquad
    \opdecl{\kode{read}}{\one}{S},
  \end{equation*}
  where $S$ is the set of entities that are read or written, for example bytes,
  or strings. There are no equations. We may now write the obligatory hello world:
  \begin{equation*}
    \opcall{print}{\text{`Hello world!'}}{\lam{\_} \return{\unit}}.
  \end{equation*}
\end{example}

\begin{example}
  \label{ex:theory-exception}
  The theory of a point set, cf.\ Example~\ref{ex:pointed-set}, is the theory of
  an \emph{exception}. The point $\bullet$ is a constant, which we rename to a
  nullary operation
  \begin{equation*}
    \opdecl{\kode{abort}}{\one}{\emptyset}.
  \end{equation*}
  There are no equations. For example, the computation
  \begin{equation*}
    \opcall{read}{\unit}{
      \lam{x}{
        \cond{x < 0}{\opcall{abort}{\unit}{\mathsf{O}_\ZZ}}{\return{(x+1)}}
      }
    }
  \end{equation*}
  reads an integer~$x$ from standard input, raises an exception if $x$ is
  negative, otherwise it returns its successor.
\end{example}

\begin{example}
  \label{ex:non-determinism}
  Let us take the theory of semilattice, cf.\ Example~\ref{ex:semi-lattice}, but without the
  unit. It has a binary operation~$\vee$ satisfying
  \begin{align*}
    x \vee x &= x, \\
    x \vee y &= y \vee x, \\
    (x \vee y) \vee z &= x \vee (y \vee z).
  \end{align*}
  This is the algebraic theory of (one variant of) \emph{non-determinism}.
  Indeed, the binary operation $\vee$ corresponds to a choice operation
  \begin{equation*}
    \opdecl{\kode{choose}}{\one}{\bool}
  \end{equation*}
  which (non-deterministically) returns a bit, or chooses a computation, depending
  on how we look at it. Written in continuation notation, it chooses a bit~$b$
  and passes it to the continuation~$\kappa$,
  \begin{equation*}
    \opcall{choose}{\unit}{\lam{b}{\kappa \, b}},
  \end{equation*}
  whereas with the traditional notation it chooses between two computations
  $\kappa_1$ and $\kappa_2$,
  \begin{equation*}
    \kode{choose}(\kappa_1, \kappa_2).
  \end{equation*}
\end{example}

\begin{example}
  \label{ex:theory-single-state}
  Algebraic theories may be combined. For example, if we want a theory
  describing state and I/O we may simply adjoin the signatures and equations of
  both theories to obtain their combination.

  Sometimes we want to combine theories so that the operations between them
  interact. To demonstrate this, let us consider the theory of a single stateful
  memory location holding elements of a set $S$. The operations are
  \begin{equation*}
    \opdecl{\kode{get}}{\one}{S}
    \qquad\text{and}\qquad
    \opdecl{\kode{put}}{S}{\one}.
  \end{equation*}
  The equations are
  \begin{align}
    \label{eq:state-get-get}%
    \opcall{get}{\unit}{
      \lam{s}{
        \opcall{get}{\unit}{
        \lam{t} \kappa \, s \, t}
      }
    } &=
    \opcall{get}{\unit}{\lam{s}{\kappa \, s \, s}}
    \\
    \label{eq:state-get-put}%
    \opcall{get}{\unit}{
      \lam{s} \opcall{put}{s}{\kappa}
    } &=
    \kappa \, ()
    \\
    \label{eq:state-put-get}%
    \opcall{put}{s}{
      \lam{\_} \opcall{get}{\unit}{\kappa}
    } &=
    \opcall{put}{s}{\lam{\_} \kappa \, s}
    \\
    \label{eq:state-put-put}%
    \opcall{put}{s}{
      \lam{\_} \opcall{put}{t}{\kappa}
    } &=
    \opcall{put}{t}{\kappa}
  \end{align}
  This is just the first group of equations from Example~\ref{ex:theory-state},
  except that we need not specify which memory location to read from. 

  Can the theory of states with many locations from
  Example~\ref{ex:theory-state} be obtained by a combination of many
  instances of the theory of a single state? That is, to model $I$-many states,
  we combine $I$-many copies of the theory of a single state, so that for every
  $\iota \in I$ we have operations
  \begin{equation*}
    \opdecl{\kode{get}_\iota}{\one}{S_\iota}
    \qquad\text{and}\qquad
    \opdecl{\kode{put}_\iota}{S_\iota}{\one},
  \end{equation*}
  with the above equations. We also need to postulate \emph{distributivity} laws
  expressing the fact that operations from instance~$\iota$ distribute over
  those of instance~$\iota'$, so long as $\iota \neq \iota'$:
  \begin{align*}
    \opcall{get_\iota}{\unit}{
       \lam{s} \opcall{get_{\iota'}}{\unit}{\lam{s'} \kappa \, s \, s'}
    } &=
    \opcall{get_{\iota'}}{\unit}{
       \lam{s'} \opcall{get_\iota}{\unit}{\lam{s} \kappa \, s \, s'}
    }
    \\
    \opcall{put_\iota}{s}{
       \lam{\_} \opcall{get_{\iota'}}{\unit}{\kappa}
    } &=
    \opcall{get_{\iota'}}{\unit}{
       \lam{t} \opcall{put_\iota}{s}{
          \lam{\_} \kappa \, t
       }
    } \\
    \opcall{put_\iota}{s}{
       \lam{\_} \opcall{update_{\iota'}}{s'}{\kappa}
    } &=
    \opcall{update_{\iota'}}{s'}{
       \lam{\_} \opcall{put_\iota}{s}{\kappa}
    }.
  \end{align*}
  The theory so obtained is similar to that of Example~\ref{ex:theory-state},
  with two important differences. First, the locations $\ell \in L$ are
  parameters of operations in Example~\ref{ex:theory-state}, whereas in the
  present case the instances $\iota \in I$ index the operations themselves.
  Second, all memory locations in Example~\ref{ex:theory-state} share the same
  set of states~$S$, whereas the combination of $I$-many separate states allows
  a different set of states $S_\iota$ for every instance~$\iota \in I$.
\end{example}

\subsection{Computations are free models}
\label{sec:comp-are-free}

Among all the models of an algebraic theory of computational effects, which one
best described the actual computational effects? If a theory of computational
effects truly is adequately described by its signature and equations, then the
free model ought to be the desired one.

\begin{example}
  Consider the theory $\theory{State}$ of a state storing elements of~$S$ from
  Example~\ref{ex:theory-single-state}. Let us verify whether the free model
  $\Free{State}{V}$ adequately describes stateful computations returning values
  from~$V$. As we saw in Section~\ref{sec:free-models}, the free model is a
  quotient of the set of trees $\Tree{\signature{State}}{V}$ by a congruence
  relation~$\approx_{\theory{State}}$. Every tree is congruent to one of the
  form
  \begin{equation}
    \label{eq:state-normal-form}
    \opcall{get}{\unit}{
      \lam{s} \opcall{put}{f(s)}{\lam{\_} \return{g(s)}}
    }
  \end{equation}
  for some maps $f : S \to S$ and $g : S \to V$. Indeed, by applying the
  equations from Example~\ref{ex:theory-single-state}, we may contract any two
  consecutive $\kode{get}$'s to a single one, and similarly for consecutive
  $\kode{put}$'s, we may disregard a $\kode{get}$ after a $\kode{put}$, and
  cancel a~$\kode{get}$ followed by a~$\kode{put}$. We are left with four
  forms of trees,
  \begin{gather*}
    \return{v},\\
    \opcall{get}{\unit}{\lam{s} \return{g(s)}},\\
    \opcall{put}{t}{\lam{\_} \return{v}},\\
    \opcall{get}{\unit}{
      \lam{s} \opcall{put}{f(s)}{\lam{\_} \return{g(s)}}
    },
  \end{gather*}
  but the first three may be brought into the form of the fourth one:
  \begin{align*}
    \return{v} &= 
    \opcall{get}{\unit}{\lam{s} \opcall{put}{s}{\lam{\_} \return{v}}},
    \\
    \opcall{get}{\unit}{\lam{s} \return{g(s)}} &=
    \opcall{get}{\unit}{\lam{s} \opcall{put}{s}{\lam{\_} \return{g(s)}}},
    \\
    \opcall{put}{t}{\lam{\_} \return{v}} &=
    \opcall{get}{\unit}{\lam{\_} \opcall{put}{t}{\lam{\_} \return{v}}}.
  \end{align*}
  Therefore, the free model~$\Free{Sate}(V)$ is isomorphic to the set of
  functions
  \begin{equation*}
    S \to S \times V.
  \end{equation*}
  The isomorphism takes the element represented by~\eqref{eq:state-normal-form}
  to the function $\lam{s} (f(s), g(s))$. (It takes extra effort to show that
  each element is represented by unique $f$ and $g$.) The inverse takes a
  function $h : S \to S \times V$ to the computation represented by the tree
  \begin{equation*}
    \opcall{get}{\unit}{
      \lam{s} \opcall{put}{\pi_1(h(s))}{\lam{\_} \return{\pi_2(h(s))}}
    }.
  \end{equation*}
  Functional programers will surely recognize the genesis of the state monad.
\end{example}

Let us expand on the last thought of the previous example and show, at the risk
of wading a bit deeper into category theory, that free models of an algebraic
theory~$\theory{T}$ form a monad. We describe the monad structure in the form of
a Kleisli triple, because it is familiar to functional programmers. First, we
have an endofunctor $\FreeFun{T}$ on the category of sets which takes a set~$X$
to the free model $\Free{T}{X}$ and a map $f : X \to Y$ to the unique
$\theory{T}$-homomorphism $\overline{f}$ for which the following diagram
commutes:
\begin{equation*}
  \xymatrix{
    {X}
    \ar[r]^{\eta_X}
    \ar[d]_{f}
    &
    **[r]{\Free{T}{X}}
    \ar[d]^{\overline{f}}
    \\
    {Y}
    \ar[r]_{\eta_Y}
    &
    **[r]{\Free{T}{Y}}
  }
\end{equation*}
Second, the unit of the monad is the map $\eta_X : X \to \Free{T}{X}$ taking~$x$
to $\return{x}$. Third, a map $\phi : X \to \Free{T}{Y}$ is lifted to the unique
map $\lift{\phi} : \Free{T}{X} \to \Free{T}{Y}$ for which the following diagram
commutes:
\begin{equation*}
  \xymatrix{
    {X}
    \ar[r]^{\eta_X}
    \ar[rd]_{\phi}
    &
    **[r]{\Free{T}{X}}
    \ar[d]^{\lift{\phi}}
    \\
    &
    **[r]{\Free{T}{Y}}
  }
\end{equation*}
Concretely, $\lift{\phi}$ is defined by recursion on
($\approx_{\theory{T}}$-equivalence classes of) trees by
\begin{align*}
  \lift{\phi}([\return{x}]) &= \phi(x), \\
  \lift{\phi}([\opcall{op}{p}{\kappa}]) &= [\opcall{op}{p}{\lift{\phi} \circ \kappa}],
\end{align*}
where~$\kode{op}$ ranges over the operations of~$\theory{T}$. The first equation
holds because the above diagram commutes, and the second because $\lift{\phi}$ is a
$\theory{T}$-homomorphism. We leave the verification of the monad laws as
an exercise.

\begin{example}
  Let us resume the previous example. If there is any beauty in mathematics, the
  monad for $\FreeFun{State}$ should be isomorphic to the usual state monad
  $(T, \theta, {}^{*})$, given by
  \begin{align*}
    T(X) &= (S \to S \times X), \\
    \theta_X(x) &= (\lam{s} (s, x)), \\
    \psi^{*}(h) &= (\lam{s} \psi (\pi_2 (h(s))) (\pi_1 (h(s)))),
  \end{align*}
  where $x \in X$, $\psi : X \to T(Y)$, and $h : S \to S \times X$. In the
  previous example we already verified that $\FreeFun{State}(X) \cong T(X)$ by
  the isomorphism
  \begin{equation*}
    \Xi :
    [\opcall{get}{\unit}{\lam{s} \opcall{put}{f(s)}{\lam{\_} \return{g(s)}}}]
    \mapsto
    (\lam{s} (f(s), g(s))).
  \end{equation*}
  Checking that~$\Xi$ transfers $\eta$ to $\theta$ and
  $\lift{{}}$ to ${}^{*}$ requires a tedious but straightforward calculation
  which is best done in the privacy of one's notebook. Nevertheless, here it is.
  Note that
  \begin{equation*}
    \eta_X(x) = [\return{x}] = [\opcall{get}{\unit}{\lam{s} \opcall{put}{s}{\lam{\_} \return{x}}}]
  \end{equation*}
  hence $\eta_X(x)$ is isomorphic to the map $x \mapsto (\lam{s} (s, x))$, which is
  just $\theta_X(x)$, as required. For lifting, consider any $\phi : X \to \Free{State}{Y}$.
  There corresponds to it a unique map $\psi : X \to (S \to S \times Y)$ satisfying
  \begin{equation*}
    \phi(x) = [\opcall{get}{\unit}{\lam{t} \opcall{put}{\pi_1(\psi(x)(t))}{\lam{\_} \return{\pi_2(\psi(x)(t)}}}].
  \end{equation*}
  First we compute $\lift{\phi}$ applied to an arbitrary element of the free model:
  \begin{align*}
    \lift{\phi}&([\opcall{get}{\unit}{\lam{s} \opcall{put}{f(s)}{\lam{\_} \return{g(s)}}}]) = \\
    &[\opcall{get}{\unit}{\lam{s} \opcall{put}{f(s)}{\lam{\_} \phi(g(s))}}] = \\
    &[\opcall{get}{\unit}{
      \lam{s}
      \opcall{put}{f(s)}{
        \lam{\_}
        \opcall{get}{\unit}{\lam{t} \opcall{put}{\pi_1 (\psi(g(s))(t))}{\lam{\_} \return{\pi_2 (\psi(g(s))(t))}}}
      }
     }] = \\
    &[\opcall{get}{\unit}{\lam{s} \opcall{put}{f(s)}{\lam{\_}
      \opcall{put}{\pi_1 (\psi(g(s))(f(s)))}{\lam{\_} \return{\pi_2 (\psi(g(s))(f(s)))}}
    }}] = \\
    &[\opcall{get}{\unit}{\lam{s}
      \opcall{put}{\pi_1 (\psi(g(s))(f(s)))}{\lam{\_} \return{\pi_2 (\psi(g(s))(f(s)))}}
    }].
  \end{align*}
  Then we compute $\psi^{*}$ applied to the corresponding element of the state monad:
  \begin{align*}
    \psi^{*}(\lam{s} (f(s), g(s)))
    &= (\lam{s} \psi(g(s))(f(s))) \\
    &= (\lam{s} (\pi_1 (\psi(g(s))(f(s))), \pi_2 (\psi(g(s))(f(s))))),
  \end{align*}
  And we have a match with respect to~$\Xi$.
\end{example}

\subsection{Sequencing and generic operations}
\label{sec:sequ-gener-oper}

We seem to have a good theory of computations, but our notation is an
abomination which neither mathematicians nor programmers would ever want to use.
Let us provide a better syntax that will make half of them happy.

Consider an algebraic theory~$\theory{T}$. For an operation $\opdecl{\kode{op}}{P}{A}$
in $\signature{T}$, define the corresponding \emph{generic operation}
\begin{equation*}
  \opgen{op}{p} \defeq \opcall{op}{p}{\lam{x} \return{x}}.
\end{equation*}
In words, the generic version performs the operation and returns its result.
When the parameter is the unit we write $\opgen{op}{}$ instead of the silly
looking $\opgen{op}{\unit}$. After a while one also grows tired of the over-line
and simplifies the notation to just $\kode{op}(p)$, but we shall not do so here.

Next, we give ourselves a better notation for the monad lifting. Suppose
$t \in \Free{T}{X}$ and $h : X \to \Free{T}{Y}$. Define the \emph{sequencing}
\begin{equation*}
  \seq{x}{t} h(x),
\end{equation*}
to be an abbreviation for $\lift{h}(t)$, with the proviso that $x$ is bound in
$h(x)$. Generic operations and sequencing allow us to replace the awkward
looking
\begin{equation*}
  \opcall{op}{p}{\lam{x} t(x)}
\end{equation*}
with
\begin{equation*}
  \seq{x}{\opgen{op}{p}} t(x).
\end{equation*}
Even better, nested operations
\begin{equation*}
  \opcall{op_1}{p_1}{\lam{x_1}
  \opcall{op_2}{p_2}{\lam{x_2}
  \opcall{op_3}{p_2}{\lam{x_3}
  \cdots
  }}}
\end{equation*}
may be written in Haskell-like notation
\begin{align*}
  &\seq{x_1}{\xopgen{\kode{op}_1}{p_1}} \\
  &\seq{x_2}{\xopgen{\kode{op}_3}{p_3}} \\
  &\seq{x_2}{\xopgen{\kode{op}_3}{p_3}}
  \cdots
\end{align*}
The syntax of a typical programming language only ever exposes the generic
operations. The generic operation $\overline{\kode{op}}$ with parameter set~$P$
and arity~$A$ looks to a programmer like a function of type $P \to A$, which is
why we use the notation $\opdecl{\kode{op}}{P}{A}$ to specify signatures.

Because sequencing is just lifting in disguise, it is governed by the same
equations as lifting:
\begin{align*}
  (\seq{x}{\return{v}} h(x)) &= h(v), \\
  (\seq{x}{\opcall{op}{p}{\kappa}} h(x)) &=
  \opcall{op}{p}{\lam{y} \seq{x}{\kappa(y)} h(x)}.
\end{align*}
These allow us to eliminate sequencing from any expression. When we rewrite the
second equation with generic operations we get an associativity law for
sequencing:
\begin{equation*}
  (\seq{x}{(\seq{y}{\opgen{op}{p}} \kappa(y))} h(x)) =
  (\seq{y}{\opgen{op}{p}} \seq{x}{\kappa(y)} h(x)).
\end{equation*}

The ML aficionados may be pleased to learn that the sequencing notation in
an ML-style language is none other than $\kode{let}$-binding,
\begin{equation*}
  \kode{let}\; x = t\;\kode{in}\;h(x).
\end{equation*}

\section{Handlers}
\label{sec:handlers}

So far the main take-away is that computations returning values from~$V$ and
performing operations of a theory~$\theory{T}$ are the elements of the free
model $\Free{T}{V}$. What about \emph{transformations} between computations,
what are they? An easy but useless answer is that they are just maps between the
carriers of free models,
\begin{equation*}
  \carrier{\Free{T}{X}} \longrightarrow \carrier{\Free{T'}{X'}},
\end{equation*}
whereas a better answer should take into account the algebraic structure. Having
put so much faith in algebra, let us continue to do so and postulate that a
transformation between computations be a homomorphism. Should it be a
homomorphism with respect to~$\theory{T}$ or~$\theory{T}'$? We could weasel out
of the question by considering only homomorphisms of the form
$\Free{T}{X} \to \Free{T}{X'}$, but such homomorphisms are rather uninteresting,
because they amount to maps~$X \to \Free{T}{X'}$. We want
transformation between computations that transform the operations as well as
values.

To get a reasonable notion of transformation, let us recall that the universal
property of free models speaks about maps \emph{from} a free model. Thus, a
transformation between computations should be a $\theory{T}$-homomorphism
\begin{equation*}
  H : \carrier{\Free{T}{X}} \longrightarrow \carrier{\Free{T'}{X'}}.
\end{equation*}
For this to make any sense, the carrier $\carrier{\Free{T'}{X'}}$ must carry the
structure of a $\theory{T}$-model, i.e., in addition to~$H$ we must also provide
a $\theory{T}$-model on $\carrier{\Free{T'}{X'}}$. If we take into account the
fact that $H$ is uniquely determined by its action on the generators, we arrive
at the following notion. A \emph{handler} from computations $\Free{T}{X}$ to
computations $\Free{T'}{X'}$ is given by the following data:
\begin{enumerate}
\item a map $f : X \to \carrier{\Free{T'}{X'}}$,
\item for every operation $\opdecl{\op{i}}{P_i}{A_i}$ in $\signature{T}$, a map
  \begin{equation*}
    h_i : P_i \times \carrier{\Free{T'}{X'}}^{A_i} \to \carrier{\Free{T'}{X'}}
  \end{equation*}
  such that
\item the maps $h_i$ form a $\theory{T}$-model on~$\carrier{\Free{T'}{X'}}$, i.e., they
 validate the equations~$\equations{T}$.
\end{enumerate}
The map $H : \carrier{\Free{T}{X}} \longrightarrow \carrier{\Free{T'}{X'}}$ induced by these
data is the unique one satisfying
\begin{align*}
  H([\return{x}]) &= f(x), \\
  H([\opcall{op}{p}{\kappa}]) &= h_i(p, H \circ \kappa).
\end{align*}
When $H$ is a handler from $\Free{T}{X}$ to $\Free{T'}{X'}$ we write
\begin{equation*}
  H : \Free{T}{X} \hto \Free{T'}{X'}.
\end{equation*}

From a mathematical point of view handlers are just a curious
combination of algebraic notions, but they are much more interesting from a
programming point of view, as practice has shown.

We need a notation for handlers that neatly collects its defining data. Let us write
\begin{equation}
  \label{eq:handler-notation}
  \handler \{
    \retclause{x} f(x), \;
    \big( \opclause{\op{i}}{y}{\kappa} h_i(y, \kappa) \big)_{\op{i} \in \signature{T}}
  \}
\end{equation}
for the handler~$H$ determined by the maps $f$ and $h_i$, as above, and
\begin{equation*}
  \withhandle{H}{C}
\end{equation*}
for the application of~$H$ to a computation~$C$. The defining equations for
handlers written in the new notation are, where $H$ stands for the
handler~\eqref{eq:handler-notation}:
\begin{align*}
  (\withhandle{H}{\return v}) &= f(v), \\
  (\withhandle{H}{\seq{x}{\xopgen{\op{i}}{p}} \kappa(x)}) &=
  h_i (p, \lam{x} \withhandle{H}{\kappa(x)})
\end{align*}

\begin{example}
  Let us consider the theory $\theory{Exn}$ of an exception, cf.\
  Example~\ref{ex:theory-exception}. A handler
  \begin{equation*}
    H : \Free{Exn}{X} \hto \Free{T}{Y}
  \end{equation*}
  is given by a $\kode{return}$ clause and an $\kode{abort}$ clause,
  \begin{equation*}
    \handler \{
      \retclause{x} f(x),
      \opclause{\kode{abort}}{y}{\kappa} c
    \},
  \end{equation*}
  where $f : X \to \Free{T}{Y}$ and $c \in \Free{T}{Y}$. Note that $c$ does not
  depend on $y \in \one$ and $\kappa : \emptyset \to \Free{T}{Y}$ because they
  are both useless. The theory of an exception has no equations, so the
  handler is automatically well defined. Such a handler is quite similar to
  exception handlers from mainstream programming languages, except that it
  handles both the exception and the return value.
\end{example}

\section{What is coalgebraic about algebraic effects and handlers?}
\label{sec:what-coalg-about}

Handlers are a form of flow control (like loops, conditional statements,
exceptions, coroutines, and the dreaded ``goto''), to be used by programmers in
programs. In other words, they can be used to \emph{simulate} computational
effects, a bit like monads can simulate computational effects in a purely
functional language. What we still lack is a mathematical model of computational
effects at the level of the external environment in which the program runs. There is
always a barrier between the program and its external environment, be it a
virtual machine, the operating system, or the underlying hardware. The actual
computational effects cross the barrier, and cannot be modeled as handlers. A
handler gets access to the continuation, but when a real computational effects
happens, the continuation is not available. If it were, then after having
launched missiles, the program could change its mind, restart the continuation,
and establish world peace.

\subsection{Comodels of algebraic theories}
\label{sec:comod-algebr-theor}

We shall model the top-level computational effects with comodels, which were
proposed by Gordon Plotkin and John
Power~\cite{plotkin08:_tensor_comod_model_operat_seman}. A \emph{comodel} of a
theory~$\theory{T}$ in a category~$\category{C}$ is a model in the opposite
category~$\opcat{\category{C}}$. Comodels form a category
\begin{equation*}
  \ComodC{\category{C}}{T} \defeq \opcat{(\ModC{\opcat{\category{C}}}{T})}.
\end{equation*}
We steer away from category-theory and just spell out what a comodel is in the
category of sets. When we pass to the dual category all morphisms turn around,
and concepts are replaced with their duals.

Recall that the interpretation of an operation $\opdecl{\kode{op}}{P}{A}$ in a
model~$M$ is a map
\begin{equation*}
  \sem{\kode{op}}_M :
  P \times |M|^A \longrightarrow |M|,
\end{equation*}
which in the curried form is
\begin{equation*}
  |M|^A \longrightarrow |M|^P.
\end{equation*}
In the opposite category the map turns its direction, and the exponentials
become products:
\begin{equation*}
  A \times |M| \longleftarrow P \times |M|.
\end{equation*}
Thus, in a comodel~$W$ an operation $\opdecl{\kode{op}}{P}{A}$ is interpreted as a map
\begin{equation*}
  \sem{\kode{op}}^W : P \times |W| \to A \times |W|,
\end{equation*}
which we call a \emph{cooperation}.

\begin{example}
  Non-deterministic choice $\opdecl{\kode{choose}}{\one}{\bool}$, cf.\
  Example~\ref{ex:non-determinism}, is interpreted as a cooperation
  \begin{equation*}
    |W| \to {\bool} \times |W|,
  \end{equation*}
  where on the left we replaced $\one \times |W|$ with the isomorphic set $|W|$.
  If we think of $|W|$ as the set of all possible worlds, the cooperation
  $\kode{choose}$ is the action by which the world produces a boolean value and
  the next state of the world. Thus an external source of binary non-determinism
  is a stream of booleans.
\end{example}

\begin{example}
  Printing to standard output $\opdecl{\kode{print}}{S}{\one}$ is interpreted as
  a cooperation
  \begin{equation*}
    S \times |W| \to |W|.
  \end{equation*}
  It is the action by which the world is modified according to the printed
  message (for example, the implants on your retina might induce your visual
  center to see the message).
\end{example}

\begin{example}
  Reading from standard input $\opdecl{\kode{read}}{\one}{S}$ is interpreted as
  a cooperation
  \begin{equation*}
    |W| \to S \times |W|.
  \end{equation*}
  This is quite similar to non-deterministic choice, except that the world
  provides an element of~$S$ rather than a boolean value. The world might
  accomplish such a task by inducing the user (who is considered as part of the
  world) to press buttons on the keyboard.
\end{example}

\begin{example}
  An exception $\opdecl{\kode{abort}}{\one}{\emptyset}$ is interpreted as a
  cooperation
  \begin{equation*}
    \one \times |W| \to \emptyset \times |W|.
  \end{equation*}
  Unless $|W|$ is the empty set, there is no such map. An exception cannot
  propagate to the outer world. The universe is safe from segmentation fault!
\end{example}

The examples are encouraging, so let us backtrack and spell out the basic
definitions properly. A \emph{cointerpretation}~$I$ of a signature $\Sigma$ is
given by a carrier set $\carrier{I}$, and for each operation symbol $\opdecl{\kode{op}}{P}{A}$
a map
\begin{equation*}
  \sem{\kode{op}}^I : P \times \carrier{I} \to A \times \carrier{I},
\end{equation*}
called a \emph{cooperation}. The cointerpretation $I$ may be extended to
well-founded trees. A tree $t \in \Tree{\Sigma}{X}$ is interpreted as a map
\begin{equation*}
  \sem{t}^I : \carrier{I} \to X \times \carrier{I}
\end{equation*}
as follows:
\begin{enumerate}
\item the tree $\leaf{x}$ is interpreted as the $x$-th injection,
  \begin{align*}
    \sem{X \mid x}^I &: \carrier{I} \to X \times \carrier{I},\\
    \sem{X \mid x}^I &: \omega \to (x, \omega).
  \end{align*}
\item the tree $\op{i}(p, \kappa)$ is interpreted as the map
  \begin{align*}
    \sem{X \mid \op{i}(p, \kappa)}^I &: \carrier{I} \to X \times \carrier{I},\\
    \sem{X \mid \op{i}(p, \kappa)}^I &: \omega \to
         \sem{X \mid \kappa(a)}^I(\varpi)
         \quad\text{where $(a, \varpi) = \sem{\op{i}}^{I}(p, \omega)$.}
  \end{align*}
\end{enumerate}
A \emph{comodel~$W$} of a theory~$\theory{T}$ is a
$\signature{T}$-cointerpretation which validates all the
equations~$\equations{T}$. As before, an equation is valid when the
interpretations of its left- and right-hand sides yield equal maps.

\begin{example}
  Let us work out what constitutes a comodel~$W$ of the theory of state, cf.\
  Example~\ref{ex:theory-single-state}. The operations
  \begin{equation*}
    \opdecl{\kode{get}}{\one}{S}
    \qquad\text{and}\qquad
    \opdecl{\kode{put}}{S}{\one}
  \end{equation*}
  are respectively interpreted by cooperations
  \begin{equation*}
    g : \carrier{W} \to S \times \carrier{W}
    \qquad\text{and}\qquad
    p : S \times \carrier{W} \to \carrier{W},
  \end{equation*}
  where we replaced $\one \times \carrier{W}$ with the isomorphic
  set~$\carrier{W}$ (and we shall continue doing so in the rest of the example).
  The cooperations~$p$ and~$g$ must satisfy the equations from
  Example~\ref{ex:theory-single-state}. We first unravel the interpretation of
  equation~\eqref{eq:state-get-put}. Recall that $\kappa$ is the generic
  continuation $\kappa \, \unit = \leaf{\unit}$, and the context contains no variables
  so it is interpreted by~$\one$. Thus the right- and left-hand sides are
  interpreted as maps $\carrier{W} \to \carrier{W}$, namely
  \begin{align*}
    \sem{\kappa\,\unit}^W &: w \mapsto w,\\
    \sem{\opcall{get}{\unit}{\lam{s} \opcall{put}{s}{\kappa}}}^W &: w \mapsto p(g(w)).
  \end{align*}
  These are equal precisely when, for all $w \in |W|$,
  \begin{equation}
    \label{eq:state-comodel-p-g}
    p(g(w)) = w.
  \end{equation}
  Keeping in mind that the dual nature of cooperations requires reading of
  expressions from inside out, so that in $p(g(w))$ the cooperation~$g$ happens
  before~$p$, the intuitive meaning of~\eqref{eq:state-comodel-p-g} is clear:
  the external world does not change when we read the state and write it right
  back. Equations~\eqref{eq:state-get-get}, \eqref{eq:state-put-get}, and
  \eqref{eq:state-put-put} may be similarly treated to respectively give
  \begin{align}
    \label{eq:state-comodel-g-g}
    g(\pi_2 (g (w))) &= g(w),
    \\
    \notag
    g(p(s, w)) &= (s, p (s, w)),
    \\
    \notag
    p(t, p(s, w)) &= p(t, w).
  \end{align}
  %
  %
  %
  %
  %
  From these equations various others can be derived. For instance,
  by~\eqref{eq:state-comodel-p-g}, the cooperation~$g$ is a section of~$p$,
  therefore we may cancel it on both sides of~\eqref{eq:state-comodel-g-g} to
  derive $\pi_2(g(w)) = w$, which says that reading the state does not alter the
  external world.
\end{example}

\begin{example}
  A comodel~$W$ of the theory of non-determinism, cf.\
  Example~\ref{ex:non-determinism}, is given by a cooperation
  \begin{equation*}
    c : \carrier{W} \to \bool \times \carrier{W}.
  \end{equation*}
  The cooperation must satisfy (the interpretations of) associativity,
  idempotency, and commutativity. Commutativity is problematic because we get
  from it that if $c(w) = (b, w')$ then also
  $c(w) = (\mathop{\mathsf{not}} b, w')$, implying the nonsensical requirement
  $b = \mathop{\mathsf{not}} b$. It appears that comodels of non-determinism require
  fancier categories than the good old sets.
\end{example}

\subsection{Tensoring comodels and models}
\label{sec:tens-comod-models}

If we construe the elements of a $\theory{T}$-model~$\carrier{M}$ as effectful
computations and the elements of a $\theory{T}$-comodel ~$\carrier{W}$ as
external environments, it is natural to ask whether $M$ and $W$ interact to give
an account of running effectful programs in effectful external environments.
Let $\sim_\theory{T}$ be the least equivalence relation
on~$\carrier{M} \times \carrier{W}$ such that, for every operation symbol
$\opdecl{\kode{op}}{P}{A}$ in $\signature{T}$, and for all $p \in P$, $a \in A$,
$\kappa : A \to \carrier{M}$, and $w, w' \in \carrier{M}$ such that
$\sem{\kode{op}}^W(p, w) = (a, w')$,
\begin{equation}
  \label{eq:tensor-equivalence}
  (\sem{\kode{op}}_M(p, \kappa), w) \sim_\theory{T} (\kappa(a), w').
\end{equation}
Define the \emph{tensor $\tensor{M}{W}$} to be the quotient set
$(\carrier{M} \times \carrier{W})/{\sim_\theory{T}}$.

The tensor represents the interaction of~$M$ and $W$. The equivalence
$\sim_\theory{T}$ in~\eqref{eq:tensor-equivalence} has an operational reading:
to perform the operation $\sem{\kode{op}}_M(p, \kappa)$ in the external
environment~$w$, run the corresponding cooperation $\sem{\kode{op}}^W(p, w)$ to
obtain $a \in A$ and a new environment~$w'$, then proceed by executing
$\kappa(a)$ in environment~$w'$.

\begin{example}
  Let us compute the tensor of $M = \Free{\theory{State}}{X}$, the free model of
  the theory of state generated by~$X$, and the comodel $W$ defined by
  \begin{equation*}
    |W| \defeq S,
    \qquad
    \sem{\kode{get}}^W \defeq \lam{s} (s, s),
    \qquad
    \sem{\kode{put}}^W \defeq \lam{(s,t)} s.
  \end{equation*}
  We may read the equivalences
  \begin{align*}
    (\sem{\opcall{get}{\unit}{\kappa}}_M, s) &\sim_{\theory{State}} (\sem{\kappa}_M (s), s), \\
    (\sem{\opcall{put}{t}{\kappa}}_M, s) &\sim_{\theory{State}} (\sem{\kappa}_M \unit, t),
  \end{align*}
  from left to right as rewrite rules which allows us to ``execute away'' all
  the operations until we are left with a pair of the form
  $(\sem{\return{x}}_M, s)$. Because
  $(\sem{\return{x}}_M, s) \sim_{\theory{State}} (\sem{\return{y}}_M, t)$
  implies $x = y$ and $s = t$ (the proof of which we skip), it follows that
  $\tensor{M}{W}$ is isomorphic to $X \times S$. In other words, the execution
  of a program in an initial state always leads to a return value paired with
  the final state.
\end{example}

The next time the subject of tensor products comes up, you may impress your
mathematician friends by mentioning that you know how to tensor software with
hardware.

\section{Making a programming language}
\label{sec:making-progr-lang}

The mathematical theory of algebraic effects and handlers may be used in
programming language design, both as a mathematical foundation and a source of
inspiration for new programming concepts. This is a broad topic which far
exceeds the purpose and scope of these notes, so we only touch on the main
questions and issues, and provide references for further reading.

\begin{figure}[ht]
  \small
  \fbox{\parbox{\textwidth}{
    \centering
    \newcommand{\bnfis}{\mathrel{\;{:}{:}\!=}\;}
    \newcommand{\bnfor}{\mathrel{\;\big|\;}}
    \begin{align*}
      \text{Value}\ v
        \bnfis& x                                & & \text{variable} \\
        \bnfor& \kode{false} \bnfor \kode{true}  & & \text{boolean constant} \\
        \bnfor& \lam{x} c                        & & \text{function} \\
        \bnfor&
          \handler \{
          \begin{aligned}[t]
            & \retclause{x} c_r, \\
            & \ldots, \opclause{\op{i}}{x}{k} c_i, \ldots \}
          \end{aligned}
          & & \text{handler}
      \\
      \text{Computation}\ c
        \bnfis& \return{v}                       & & \text{pure computation} \\
        \bnfor& \opgen{op}{v}                    & & \text{operation} \\
        \bnfor& \seq{x}{c_1} c_2                 & & \text{sequencing} \\
        \bnfor& \cond{v}{c_1}{c_2}               & & \text{conditional} \\
        \bnfor& v_1 \, v_2                       & & \text{application} \\
        \bnfor& \withhandle{v}{c}                & & \text{handling}
      \\
      \text{Value type}\ A, B
         \bnfis& \kode{bool}                     & & \text{boolean type} \\
         \bnfor& A \times B                      & & \text{product type} \\
         \bnfor& A \to \ct{C}                    & & \text{function type} \\
         \bnfor& \ct{C} \hto \ct{D}              & & \text{handler type}
      \\
      \text{Computation type}\ \ct{C}, \ct{D}
         \bnfis& \dirt{A}{\{\op{1}, \ldots, \op{k}\}}
    \end{align*}
  }}
  \caption{A core language with algebraic effects and handlers}
  \label{fig:mini-eff}
\end{figure}

Figure~\ref{fig:mini-eff} shows the outline of a core language based on
algebraic theories, as presented so far. Apart from a couple of changes in
terminology and notation there is nothing new. Instead of generators and
generating sets we speak of \emph{values} and \emph{value types}, and instead of
trees and free models we speak of \emph{computations} and \emph{computation
  types}. The computation type $\dirt{A}{\{\op{1}, \ldots, \op{k}\}}$
corresponds to the free model $\Free{T}{A}$ where $\theory{T}$ is the theory
with operations $\op{1}, \ldots, \op{k}$ without any equations. The rest of the
table should look familiar. And operational semantics and typing rules still
have to be given. For these we refer to Matija Pretnar's
tutorial~\cite{pretnar15:_introd_algeb_effec_handl}, and
to~\cite{bauer14:_effec_system_algeb_effec_handl,pretnar14:_infer_algeb_effec}
for a more thorough treatment of the language.

The programming language in Figure~\ref{fig:mini-eff} can express only the
terminating computations. To make it more realistic, we should add to it general
recursion and allow non-terminating computations. Such modifications cannot be
accommodated by the set-theoretic semantics, but they can be handled by domain
theory, as was shown in~\cite{bauer14:_effec_system_algeb_effec_handl}. Therein
you can find an adequate domain-theoretic semantics for algebraic effects and
handlers with support for general recursion.

Once a programming language is in place, the next task is to explore its
possibilities. Are user-defined operations and handlers good for anything?
Practice so far has shown that indeed they can be used for all sorts of things,
but also that it is possible to overuse and misuse them, just like any
programming concept. Handlers have turned out to be a versatile tool that
unifies and generalizes a number of techniques: exception handlers, backtracking
and other search strategies, I/O redirection, transactional memory, coroutines,
cooperative multi-threading, delimited continuations, probabilistic programming,
and many others. As this note is already getting quite long, we recommend
existing
material~\cite{pretnar15:_introd_algeb_effec_handl,bauer15:_progr,kammar13:_handl}
for further reading. For experimenting with handlers in practice, you can try
out one of the languages that implements handlers. The first such language was
Eff~\cite{bauer:_eff}, but there are by now others. The Effects Rosetta
Stone~\cite{effec_roset_stone} is a good starting point to learn about them and
to see how they compare. The Effect bibliography~\cite{effec} is a good source
for finding out what has been published in the area of computational effects.

\section{Exploring new territories}
\label{sec:expl-new-terr}

Lastly, we mention several aspects of algebraic effects and handlers that have
largely remained unexplored so far.

Perhaps the most obvious one is that existing implementations of effects and
handlers largely ignore equations. In a sense this is expected and
understandable. For~\eqref{eq:handler-notation} to define a handler
$\Free{T}{X} \hto \Free{T'}{Y}$, the operation clauses $h_i$ must satisfy the
equations of~$\theory{T}$. In general it is impossible to check algorithmically
whether this is the case, and so a compiler or a language interpreter should
avoid trying to do so. Thus existing languages with handlers solve the problem
by ignoring the equations. This is not as bad as it sounds, because in practice
we often want handlers that break the equations. Moreover, dropping equations
just means that we work with trees as representatives of their equivalence
classes, which is a common implementation technique (for instance, when we
represent finite sets by lists). Nevertheless, incorporating equations into
programming languages would have many benefits.

The idea of tensoring comodels and models as a mathematical explanation of the
interaction between a program and its external environment is very pleasing, but
has largely not been taken advantage of. There should be a useful programming
concept in there, especially if we can make tensoring a user-definable feature
of a programming language. The only known (to me) attempt to do so were the
\emph{resources} in an early version of Eff~\cite{bauer15:_progr}, but those
disappeared from later versions of the language. May they see the light of day
again.

At the Dagstuhl seminar~\cite{chandrasekaran18:_algeb} the topic of dynamic
creation of computational effects was recognized as important and mostly
unsolved. The operations of an algebraic theory are fixed by the signature, but
in real-world situations new instances of computational effects are created and
destroyed dynamically, for example, when a program opens or closes a file,
allocates or deallocates memory, spawns or terminates a new thread, etc. How
should such phenomena be accounted for mathematically? A good answer would
likely lead to new programming concepts for general resource management. Once
again, the only known implementation of dynamically created instances of effects
was provided in the original version of Eff~\cite{bauer15:_progr}, although some
languages allow dynamic creation of new effects by indirect means.

\bibliographystyle{plain}
\bibliography{what-is-algebraic}

\end{document}